\documentclass[12pt]{article}
\usepackage{amsbsy}
\usepackage{amssymb}
\usepackage{color}

\newcommand{\beq}{\begin{equation}}
\newcommand{\eeq}{\end{equation}}
\newcommand{\bdm}{\begin{displaymath}}
\newcommand{\edm}{\end{displaymath}}
\newcommand{\beqr}{\begin{eqnarray}}
\newcommand{\eeqr}{\end{eqnarray}}
\newcommand{\beqrn}{\begin{eqnarray*}}
\newcommand{\eeqrn}{\end{eqnarray*}}

\def\l{\lambda}
\def\bchi{\boldsymbol{\chi}}
\def\ve{\varepsilon}

\def\k{\kappa}
\def\l{\lambda}
\def\nn{\nonumber}
\def\ni{\noindent}

\begin{document}

\title{Some results on generating functions for characters and weight multiplicities of the Lie algebra $A_3$}

\author{Jos\'e Fern\'andez N\'u\~{n}ez$^{\dagger}$, 
Wifredo Garc\'{\i}a Fuertes
$^{\ddagger}$
\\
\small Departamento de F\'\i sica, 
Facultad de Ciencias, Universidad de Oviedo, 33007-Oviedo, Spain\\
\small {\it $^\dagger$nonius@uniovi.es; $^\ddagger$
wifredo@uniovi.es}\\
\and
 Askold M. Perelomov\\
\small Institute for Theoretical and Experimental Physics, Moscow, Russia.\\
\small {\it  aperelomo@gmail.com}}

\date{ }

\maketitle

\begin{abstract}
\noindent
A method based on the quantum Calogero-Sutherland model is used to obtain generating functions for characters and multiplicities of $A_3$. Some comments on other rank three algebras are offered.
\end{abstract}
\bigskip


\medskip

\vfill\eject 
\section{Introduction}
Generating functions for characters are very  useful tools for the study of representations of Lie algebras. Since they were introduced in \cite{ps79}, different procedures for computing these generating functions and extracting practical information from them have been developed, see for instance \cite{ow07} and references therein. We have recently shown how the theory of the  quantum integrable Calogero-Sutherland models can be used to obtain the generating functions \cite{nfp14} in an efficient manner, and we have obtained, proceeding on this basis, several results regarding characters and weight multiplicities for the rank two Lie algebras \cite{nfp15a,nfp15b}. Compared to other schemes, the approach based on the Calogero-Sutheland model is advantageous in that cumbersome combinatorial recipes involving the Weyl group are bypassed. This makes the method a rather convenient one to be applied to any Lie algebra, but an unavoidable fact is that, as the rank increases, the final result for the full generating function of characters become quickly exceedingly complicated. Thus, in general, for higher rank algebras only the generating functions for some restricted sets of characters, typically with only one or two non-vanishing Dynkin indices, are tractable without falling into an excessive clumsiness. Apart from rank two algebras, the only exception seems to be the case of $A_3$, in which the Weyl orbits of the fundamental weights, the Weyl formula for dimensions, and the Calogero-Sutherland Hamiltonian, are still quite simple, these circumstances suggesting that the whole generating function for characters is not too unwieldy to be computed. As far as we know, this generating function has not been explicitly written down in previous works. It seems thus worth to use the method of \cite{nfp14} to obtain it and to explore some other results which can be deduced from it. 

Results of this type are interesting in themselves, but also due to the fact that $A_3\simeq D_3$ is a Lie algebra with applications to important physical systems. Let us mention, among several others, the grand unified Pati-Salam model with gauge group $SU(2)_L \times SU(2)_R \times SU(4)$ \cite{ps74}, which has been much investigated owing to the fact that it fits well with the string or M-theory framework and it is also suitable to explain phenomenological issues like neutrino oscillations or baryogenesis \cite{jp06}; the effective $SU(4)$ hadronic models which, in spite of the large breaking of $SU(4)$ flavor symmetry, can be used to study the phenomenology of charmed particles, see for instance \cite{hglkl}; the $SU(4)$-Kondo effect, due to the interplay between spin and orbital electronic degrees of freedom, which has recently aroused remarkable interest \cite{ketal13} because of its role in condensed matter settings such as quantum dots, carbon nanotubes or nanowires; or the $AdS_5/CFT_4$ string-gauge equivalence \cite{ma98}, in which $SU(4)$ is the R-symmetry in the supersymmetric quantum field theory side of the duality and $SO(6)$ is the symmetry of the effective IIB gauged supergravity on the string side.

\section{The generating function for characters}
This section is devoted to the computation of the generating function for the characters of the irreducible representations of $A_3$ by means of the approach developed in \cite{nfp14}. This approach needs a background on Calogero-Sutherland models \cite{ca71,su72}, a class of integrable dynamical systems with interactions modeled on the root systems of Lie algebras \cite{op76}.  The background is succinctly explained in \cite{nfp14}, which also includes references to more detailed expositions of the subject.  For the case of the Lie algebra $A_3$, the model describes the classical or quantum dynamics of a system of four particles moving on a circle and interacting through a pairwise potential of trigonometric form, whose strength is governed by a single coupling constant $\kappa$. In the quantum regime the Schr\"{o}dinger equation leads to an eigenvalue problem
\beq
\Delta\Phi_{m}(q)=\varepsilon_m\,\Phi_{m}(q)\,, \label{eq:sch}
\eeq
where $q=(q_1,q_2,q_3,q_4)$ are the coordinates of the particles along the circle and $\Delta$ is a linear differential operator obtained from the original Hamiltonian. The eigenfunctions $\Phi_m$  and the eigenvalues $\ve_m$  are indexed by the 3-tuples of non-negative integers $m=(m_1,m_2,m_3)$  --the quantum numbers--,  and $m_1\l_1+m_2\l_2+m_3\l_3$  are the   highest weights of the irreducible representations of the algebra, with $\l_1,\l_2,\l_3$ being the fundamental weights. For our purposes, the relevant fact is that in the case $\k=1$ the eigenfunctions  are exactly the irreducible characters $\bchi_{m_1,m_2,m_3}$ of the algebra $A_3$.

Due to the Weyl-invariance of the Hamiltonian, it is advantageous to describe the system in terms of a set of independent W-invariant variables $z_1,z_2,z_3$, namely the characters of the three fundamental representations $R_{\l_k}$of $A_3$, which are related to the $q$-variables by  
\beqr
&&z_1=\bchi_{1,0,0}=x_1+\frac{x_2}{x_1}+\frac{x_3}{x_2}+\frac{1}{x_3}\,,\nonumber\\ 
&&z_2=\bchi_{0,1,0}=x_2 + \frac{x_3}{x_1} + \frac{x_2}{x_1 x_3} + \frac{x_1 x_3}{x_2} + \frac{x_1}{x_3} + \frac{1}{x_2}\,, \label{eq:fundchar}\\ 
&&z_3=\bchi_{0,0,1}=x_3+\frac{x_2}{x_3}+\frac{x_1}{x_2}+\frac{1}{x_1}\,,\nonumber
\eeqr
where $x_j=e^{2 {\rm i}\lambda_j\cdot q}$, the fundamental weights  $\l_j$ of $A_3$ given as four-tuples in the standard way. Higher order characters are $z$-polynomials with integer coefficients. The change of variables from the coordinates on the circle to the fundamental characters, see \cite{flp01}, leads to the following form $\Delta_{(z)}$ of the Hamiltonian $\Delta$ in terms of the $z$-variables:
\beqr
\label{eq:hz}
&&\Delta_z=\frac{1}{2}\left[(3 z_1^2-8 z_2)\partial_{z_1}^2+(4 z_2^2-8 z_1 z_3-16)\partial_{z_2}^2+(3 z_3^2-8 z_2)\partial_{z_3}^2+(4z_1 z_2-24 z_3)\partial_{z_1}\partial_{z_2}\right.  \nn            \\&&\qquad\ +\left. (2 z_1 z_3-32)\partial_{z_1}\partial_{z_3}+(4z_2 z_3-24 z_1)\partial_{z_2}\partial_{z_3}+15 z_1 \partial_{z_1}+20 z_2 \partial_{z_2}+15 z_3 \partial_{z_3}\right];
\eeqr
the eigenvalues, on the other hand, are
\beqr
\ve_m=\frac{1}{2} (3 m_1^2 + 4 m_2^2 + 3 m_3^2+ 4 m_1 m_2 + 2 m_1 m_3 + 4 m_2 m_3 + 12 m_1 + 16 m_2 + 12 m_3).\label{eq:eig}
\eeqr

In this setup, as explained in \cite{nfp14} the generating function for characters
\beq
G(t_1,t_2,t_3;z_1,z_2,z_3)=\sum_{m_1=0}^\infty \sum_{m_2=0}^\infty \sum_{m_3=0}^\infty t_1^{m_1}t_2^{m_2}t_3^{m_3}\bchi_{m_1,m_2,m_3}(z_1,z_2,z_3)\label{eq:genchar}
\eeq
is a rational function 
\beq
G(t_1,t_2,t_3;z_1,z_2,z_3)=\frac{N(t_1,t_2,t_3;z_1,z_2,z_3)}{D(t_1,t_2,t_3;z_1,z_2,z_3)} \label{eq:ga3}
\eeq
 that can be obtained by solving the differential equation
\beq
(\Delta_t-\Delta_z)G(t_k;z_k)=0, \label{eq:dif}
\eeq
where the differential operator $\Delta_t$ arises by performing the substitution $m_i\rightarrow t_i \partial_{t_i}$ in (\ref{eq:eig}). The computation goes through the following four steps: 

\medskip
\ni{\bf Step (i):} The denominator of the generating function is
\beq
D(t_1,t_2,t_3;z_1,z_2,z_3)=D_1\times D_2\times D_3 \,,\label{eq:deno}
\eeq
where $D_i=\prod_{j} (1-t_i x_1^{n_{j1}} x_2^{n_{j2}} x_3^{n_{j3}})$ with the product extended to all the weights $\sum_{k=1}^3 n_{jk} \lambda_k$ entering in the Weyl orbit of the fundamental representation $R_{\lambda_i}$. Looking at (\ref{eq:fundchar}) and bearing in mind that the fundamental representations of $A_3$ contain one single Weyl orbit, the result follows. After changing variables back from the $x_j$ to the $z_k$, it can be written as
\beqr
&&D_1=1-t_1 z_1+t_1^2 z_2-t_1^3 z_3+t_1^4\,,\nn\\
&&D_2=1+t_2^6-(t_2+t_2^5) z_2+(t_2^2+t_2^4) (z_1 z_3-1)+t_2^3 (2 z_2-z_1^2-z_3^2)\,,\label{eq:factdeno}\\
&&D_3=1-t_3 z_3+t_3^2 z_2-t_3^3 z_1+t_3^4 \,.\nn
\eeqr

\medskip
\noindent{\bf Step (ii):} The Weyl formula for dimensions gives for the representation $R_\l$, $\l=\sum_i m_i \lambda_i$,
\beq
\dim R_{\l}= \frac{1}{12} (m_1+1) (m_2+1) (m_3+1) (m_1 + m_2+2) (m_2 + m_3+2) (m_1 + m_2 + m_3+3). 
\eeq

To obtain the generating function for dimensions $E(t_1,t_2,t_3)$ it suffices to perform the change $m_i\rightarrow t_i \partial_{t_i}$ in this formula and to apply the resulting differential operator to $\prod_{i=1}^3(1-t_i)^{-1}$.  One finds in this way that
\beq
\label{eq:dim}
E(t_1,t_2,t_3)=\frac{P(t_1,t_2,t_3)}{(1-t_1)^4(1-t_2)^6 (1-t_3)^4}\,,
\eeq
where
\beqrn
P(t_1,t_2,t_3)&=&1 - 4 t_1 t_2 -4 t_2 t_3- t_1 t_3- t_2^2+ t_1^2 t_2+ t_2 t_3^2+ 4 t_1 t_2^2+ 4 t_2^2 t_3+ 6 t_1 t_2 t_3- t_1^2 t_2^3\\ &-&t_2^3 t_3^2- 4 t_1^2 t_2^2 t_3- 4 t_1 t_2^2 t_3^2- 6 t_1 t_2^3 t_3+ t_1 t_2^4 t_3+ t_1^2 t_2^2 t_3^2+ 4 t_1 t_2^3 t_3^2+ 4 t_1^2 t_2^3 t_3- t_1^2 t_2^4 t_3^2\,.
\eeqrn

A further simplification is possible, but we have written $E(t_1,t_2,t_3)$ in such a way that the denominator comes from the substitution in (\ref{eq:deno}) of the fundamental characters $z_1,z_2$ and $z_3$ by their dimensions.

\medskip
\noindent{\bf Step (iii):} We will compute the numerator $N(t_1,t_2,t_3;z_1,z_2,z_3)$ of the generating function of characters (\ref{eq:ga3}) by tentatively assuming that it contains only the terms appearing in the numerator $P(t_1,t_2,t_3)$ of (\ref{eq:dim}), now with coefficients depending of the $z$-variables. Under such hypothesis, we can expand ${N}/{D}$ as a series in the $t$-variables and compare with the right-hand side of (\ref{eq:genchar}), so that we will be able to fix the coefficients in $N(t_1,t_2,t_3;z_1,z_2,z_3)$ provided that the expressions of some low-order characters of $A_3$ are known. To obtain these is not a difficult task: as we have said, they are polynomials in the $z$-variables and, given the simple structure of the Hamiltonian $\Delta_z$ (\ref{eq:hz}), they can be computed by recursively solving the eigenvalue equation (\ref{eq:sch})\footnote{To compute the characters, or to check other results of the paper, see the ancillary files attached to the preprint.}. Thus we obtain
\beqr
N(t_1,t_2,t_3;z_1,z_2,z_3)&=&1 - z_3 t_1 t_2 -z_1 t_2 t_3- t_1 t_3- t_2^2+ t_1^2 t_2+ t_2 t_3^2+ z_1 t_1 t_2^2\nonumber\\&+& z_3 t_2^2 t_3+ z_2 t_1 t_2 t_3 -t_1^2 t_2^3- t_2^3 t_3^2- z_1 t_1^2 t_2^2 t_3- z_3 t_1 t_2^2 t_3^2\nonumber\\&-&  z_2 t_1 t_2^3 t_3+ t_1 t_2^4 t_3+ t_1^2 t_2^2 t_3^2+z_1 t_1 t_2^3 t_3^2+ z_3 t_1^2 t_2^3 t_3- t_1^2 t_2^4 t_3^2\,. \label{eq:numgen}
\eeqr

\medskip
\noindent{\bf Step (iv):} We need to be sure that the conjecture to limit the number of unknown coefficients in step (iii) is correct. For this purpose, we have to verify that $G=N/D$ (\ref{eq:ga3})
does indeed satisfy the differential equation (\ref{eq:dif}). This is a matter of directly plugging (\ref{eq:ga3}) into (\ref{eq:dif}) and doing the derivatives. In this way, one can check  that (\ref{eq:dif}) is fulfilled. Thus (\ref{eq:ga3}) with (\ref{eq:deno}), (\ref{eq:factdeno}), and (\ref{eq:numgen}) is the correct generating function for characters of the algebra $A_3$.
\section{Generating functions for weight multiplicities}
Once we have the generating function for characters, it is possible to use it to obtain some other results. Let us consider, in particular, generating functions of the form
\beq
\label{Amu}
A_{n_1,n_2,n_3}(t_1,t_2,t_3)=\sum_{m_1=0}^\infty\sum_{m_2=0}^\infty\sum_{m_3=0}^\infty\mu_{m_1,m_2,m_3}(n_1,n_2,n_3) t_1^{m_1}\, t_2^{m_2}\, t_3^{m_3}\,, 
\eeq
where $\mu_{m_1,m_2,m_3}(n_1,n_2,n_3)$ is the multiplicity of the weight $n_1\lambda_1+n_2\lambda_2+n_3\lambda_3$ in the representation $R_{m_1\lambda_1+m_2\lambda_2+m_3\lambda_3}$ of $A_3$. The way in which these generating functions can be computed is described in \cite{nfp15a,nfp15b}, see also \cite{dok}: after expressing $G(t_1,t_2,t_3;z_1,z_2,z_3)$ in  the $x$-variables by means of (\ref{eq:fundchar}), they are given by the triple integral
\beq
\label{eq:int}
A_{n_1,n_2,n_3}(t_1,t_2,t_3)=\frac{1}{(2\pi i)^3}\oint d x_3\oint d x_1\oint d x_2 \frac{G(t_1,t_2,t_3;x_1,x_2,x_3)}
{x_1^{1+n_1} x_2^{1+n_2} x_3^{1+n_3}}\,,
\eeq
where the  integration contours are along the unit circles on the complex $x_1$, $x_2$ and $x_3$-planes. We will give here explicit expressions of $A_{n_1,n_2,n_3}(t_1,t_2,t_3)$ for $n_1+n_2+n_3\leq 2$. In all these cases the integrations (\ref{eq:int}), which are readily performed by means of the residue theorem, go along the same pattern. First, the integral in $x_2$ acquires contributions from poles arising at $x_2=t_1$, $x_2=t_1 x_3$, $x_2=t_3 x_1$ and $x_2=t_2 x_1 x_3$; then, for the integral in $x_1$ there are poles at $x_1=t_3$, $x_1=t_1^2 x_3$, $x_1=t_2 x_3$ and $x_1=t_2^2 x_3^{-1}$; finally, the poles contributing to the last integral are located at $x_3=t_1$, $x_3=t_2 t_3$, $x_3=t_3^3$, $x_3=t_2^{3/2}$ and $x_3=t_2^{-3/2}$, except for the case $(n_1,n_2,n_3)=(2,0,0)$, where an additional pole at $x_3=0$ occurs. After the residues are evaluated, we find the final results for (\ref{Amu}) in the form
\bdm
A_{n_1,n_2,n_3}(t_1,t_2,t_3)=\frac{N_{n_1,n_2,n_3}(t_1,t_2,t_3)}{D_0(t_1,t_2,t_3)}\,,
\edm
where the denominator is in all cases
\bdm
D_0(t_1,t_2,t_3)= (1 - t_1^4) (1 - t_3^4) (1 - t_1 t_3)^2 (1 - t_1^2 t_2) (1 - t_2 t_3^2)(1 - t_2^2)^2
\edm
and the numerators are given in the Appendix

A possible application of the generating function $A_{n_1,n_2,n_2}(t_1,t_2,t_3)$ is  to obtain closed formulas for the multiplicities $\mu_{m_1,m_2,m_3}(n_1,n_2,n_3)$ by proceeding as done for rank two algebras in \cite{nfp15b}. Nevertheless, in the case of $A_3$ the expressions given above are somewhat complicated and the procedure turns out to be considerably cumbersome, as are indeed other approaches: see for instance \cite{ta62}, or \cite{kp14} for a recent computation of $\mu_{m_1,m_2,m_3}(0,0,0)$. We have studied the case of the real weights in the previous list by means of the Kostant multiplicity formula \cite{ko59}, see \cite{fh91} for a pedagogic exposition,
\bdm
\mu_{m_1,m_2,m_3}(n_1,n_2,n_3)=\sum_{w\in W} (-1)^w {\cal Z}\left[w(\sum_{i=1}^3 (m_i+1)\lambda_i)-\sum_{i=1}^3 (n_i+1)\lambda_i\right],
\edm
where $W$ is the Weyl group and ${\cal Z}[\sum_{i=1}^3 k_i \alpha_i]\equiv {\cal Z}[k_1,k_2,k_3]$ is the Kostant partition function for $A_3$. This function gives the number of different ways in which a vector of the root lattice can be expressed as a linear combination of the positive roots with non-negative integer coefficients. The generating function for ${\cal Z}[k_1,k_2,k_3]$ is
\bdm
\sum_{k_1=0}^\infty\sum_{k_2=0}^\infty\sum_{k_3=0}^\infty t_1^{k_1} t_2^{k_2} t_3^{k_3} {\cal Z}[k_1,k_2,k_3]=\frac{1}{(1 - t_1) (1 - t_2) (1 - t_3) (1 - t_1 t_2 t_3) (1 - t_1 t_2) (1 - t_2 t_3)}\,.
\edm
Thus, ${\cal Z}[k_1,k_2,k_3]$ is symmetric under interchange of $k_1$ and $k_3$ and its expression for $k_1\leq k_3$ can eventually found to be
\beqrn
6\, {\cal Z}[k_1,k_2,k_3] &=&(k_2+1) (k_2+2) (k_2+3)\,,\\
6\, {\cal Z}[k_1,k_2,k_3] &=& (k_1+1) (k_1+2) (3 k_2 - 2 k_1+3)\,,\\
6\, {\cal Z}[k_1,k_2,k_3] &=&(k_1+1) (k_1+2) (3 k_3 - k_1+3)\,,\\
6\,{\cal Z}[k_1,k_2,k_3] &=&(k_2 - k_3+1) (k_2 - k_3+2) (2 k_2- 3 k_1 + k_3+3)\\ &-& 
     (k_1 - k_2 + k_3) ( 3 k_1+ 3 k_3- 12 k_2 + 2 k_1^2 + 2 k_3^2- k_2^2 - k_1 k_2 - 2 k_1 k_3 - k_2 k_3 -11)\,,
\eeqrn
for, respectively, the cases i) $k_1\geq k_2$, ii) $k_1< k_2, k_3\geq k_2$, iii) $k_3< k_2, k_1 \leq k_2-k_3$ and iv)~$k_3<k_2, k_1> k_2-k_3$. With this, and taking advantage of the symmetry under $m_1\leftrightarrow m_3$ to state the results only for $m_1\leq m_3$, one finds the following formulas:
\begin{itemize}
\item  $\mu_{m_1,m_2,m_3}(0,0,0)\neq 0$ only if $m_3-m_1=2m_2+4 p$ with $p$ integer, and in this case
\beqrn
\mu_{m_1, m_2, m_3}(0,0,0)&=& a(m_1,m_2,m_3)\hspace{4cm}  {\rm if\ } p\geq 0\,,\\
8\,\mu_{m_1, m_2, m_3}(0,0,0)&=& (m_1+1)[b(m_1,m_2,m_3)+8]\hspace{1.55cm}  {\rm if\ } p<0\,;
\eeqrn
\item  $\mu_{m_1,m_2,m_3}(0,1,0)\neq 0$ only if $m_3-m_1=2(m_2-1)+4 p$ with $p$ integer, and in this case
\beqrn
\mu_{m_1, m_2, m_3}(0,1,0)&=& a(m_1,m_2,m_3)-2 (m_1+1) \delta_{p,0}\hspace{1.6cm}  {\rm if\ } p\geq 0\,,\\
8\,\mu_{m_1, m_2, m_3}(0,1,0)&=&(m_1+1)[b(m_1,m_2,m_3)+4]\hspace{2cm}  {\rm if\ } p<0\,;
\eeqrn
\item  $\mu_{m_1,m_2,m_3}(1,0,1)\neq 0$ only if $m_3-m_1=2m_2+4 p$ with $p$ integer, and in this case
\beqrn
\mu_{m_1, m_2, m_3}(1,0,1)&=& a(m_1,m_2,m_3)-(m_1+1)\delta_{p,0}\hspace{1.85cm}  {\rm if\ } p\geq 0\,,\\
8\,\mu_{m_1, m_2, m_3}(1,0,1)&=&(m_1+1) b(m_1,m_2,m_3)\hspace{3cm}  {\rm if\ } p<0\,;
\eeqrn
\item $\mu_{m_1, m_2, m_3}(0,2,0)\neq 0$ only if  $m_3-m_1=2(m_2-2)+4 p$ with $p$ integer, and in this case
\beqrn
\mu_{m_1, m_2, m_3}(0,2,0)&=& a(m_1,m_2,m_3)-2(m_1+1)(\delta_{p,1}+3\delta_{p,0})\\&+&\delta_{p,1}\delta_{m_2, 0}+\delta_{p,0}\delta_{m_2, 2}\hspace{4.5cm} {\rm if\ } p\geq 0\,,\\
8\, \mu_{m_1, m_2, m_3}(0,2,0)&=&  (m_1+1) [b(m_1,m_2,m_3)-8] +8 \delta_{m_1, m_3} \hspace{1.3cm} {\rm if\ } p<0\,,
\eeqrn
\end{itemize}
where 
\beqrn
a(m_1,m_2,m_3)&=&\frac{1}{2}(m_1 + 1)(m_2 + 1)(m_1 + m_2 + 2)\,,\\
b(m_1,m_2,m_3)&=&4\left[ (m_2+1)(m_3+1)-1\right]-(m_1-m_3)^2 .
\eeqrn

The derivation of these expressions from the Kostant multiplicity formula is a laborious process: the Weyl group of $A_3$ has order 24 and hence there are many different cases which must be separately considered and then assembled together. Thus, to give a detailed description of the proof of these results is pretty tedious. Nevertheless, once they are written down, the generating functions for multiplicities given above provide a practical way to check that they are correct. In each case, with the help of a program for symbolic computations, it is easy to expand the generating function as a Taylor series in $t$-variables  up to some high order and to subtract from this expansion the corresponding series built with the $\mu_{m_1, m_2, m_3}(n_1,n_2,n_3)$ coefficients. One then finds that the difference is zero, as it should be. 
\section{Generating function for the characters of real representations}
The generating function obtained in Section 2 collects together the characters of all irreducible representations of $A_3$. It can be of interest to have also generating functions for particular subsets of characters. The simplest examples are the generating functions for characters with only one or two non-vanishing Dynkin indices, which follow directly from (\ref{eq:ga3}) when the appropriate $t$-variables are taken to vanish. A more interesting distinction is between the characters of complex and real representations, the latter being those  with highest weight symmetric under interchange of $z_1$ and $z_3$, i.e., of the form $\bchi_{m_1,m_2,m_1}$. The general four-step procedure used in Section 2 can be also applied to construct the generating function for characters of this type,
\bdm
G_{\bf R}(t_1,t_2;z_1,z_2,z_3)=\sum_{m_1=0}^\infty \sum_{m_2=0}^\infty t_1^{m_1}t_2^{m_2}\bchi_{m_1,m_2,m_1}(z_1,z_2,z_3),
\edm
as follows:
 
 \medskip
\noindent{\bf Step (i):} Assuming that the generating function $G_{\bf R}$ is rational, the denominator  is now
\beq
D_{\bf R}(t_1,t_2;z_1,z_2,z_3)=D_{13}\times D_2 \,,\label{eq:denoreal}
\eeq
where the weights entering in $D_{13}$ are those in the Weyl orbit $R_{\lambda_1+\lambda_3}$. These can be read from the corresponding monomial symmetric function
\bdm
M_{\lambda_1+\lambda_3}=x_1 x_3 + \frac{x_1 x_2}{x_3}+\frac{ x_1^2}{x_2} + \frac{x_2 x_3}{x_1} +\frac{ x_2^2}{x_1 x_3} +\frac{x_3^2}{x_2}+{\rm c.c.}
\edm
and lead to the expression
\beqrn
D_{13}&=&1 + t_1^{12} - (t_1+t_1^{11}) (z_1 z_3-4) + 
(t_1^2+t_1^{10}) (z_1^2 z_2 - 2 z_2^2 - 4 z_1 z_3 + z_2 z_3^2+10) \\&-&
(t_1^3+t_1^9) ( z_1^4 - 7 z_1^2 z_2 + 8 z_2^2 + 13 z_1 z_3 + z_1 z_2^2 z_3 - 
    7 z_2 z_3^2 + z_3^4-20)\\&+&
(t_1^4+t_1^8)d_{48}+
(t_1^5+t_1^7)d_{57}+
 t_1^6 d_6\,,
\eeqrn
where
\beqrn
d_{48}&=& z_2^4 - 3 z_1^4 + 18 z_1^2 z_2 - 16 z_2^2  - 24 z_1 z_3 + 
    z_1^3 z_2 z_3 - 8 z_1 z_2^2 z_3 - z_1^2 z_3^2 + 18 z_2 z_3^2 + z_1 z_2 z_3^3 \\&-& 
    3 z_3^4+31\,,\\
d_{57}&=& 4 z_2^4 - 6 z_1^4 + 29 z_1^2 z_2 - 24 z_2^2 - z_1^2 z_2^3  - 
    34 z_1 z_3 + 5 z_1^3 z_2 z_3 - 19 z_1 z_2^2 z_3 - 2 z_1^2 z_3^2 + 
    29 z_2 z_3^2\\ &-& z_2^3 z_3^2 - z_1^3 z_3^3 + 5 z_1 z_2 z_3^3 - 6 z_3^4+40\,,\\
d_6&=& z_1^2 z_2^2 z_3^2 - 7 z_1^4 + 34 z_1^2 z_2 - 28 z_2^2 - 2 z_1^2 z_2^3 + 6 z_2^4 - 
    40 z_1 z_3 + 6 z_1^3 z_2 z_3 - 24 z_1 z_2^2 z_3 + 34 z_2 z_3^2  
    \\&-& 2 z_2^3 z_3^2 - 2 z_1^3 z_3^3 + 6 z_1 z_2 z_3^3 - 
    7 z_3^4+44\,.
\eeqrn

\medskip
\noindent{\bf Step (ii):} For real representations, the dimensions are
\bdm
\dim R_{m_1 \lambda_1+m_2 \lambda_2+m_1 \lambda_3}= \frac{1}{
  12} (m_1+1)^2 (m_2 + 1) (m_1 + m_2+2)^2 (2 m_1 + m_2+3) \,.
\edm
Given this formula, we can proceed as in Section 2 to  shape the generating function for dimensions. It turns out to be
\bdm
E_{\bf R}(t_1,t_2)=\frac{(1-t_1)^6(1-t_2)P_{\bf R}(t_1,t_2)}{(1 - t_1)^{12} (1 - t_2)^6},
\edm
where
\beqrn
P_{\bf R}(t_1,t_2)&=&1 + 9 t_1 + 9 t_1^2 + t_1^3 + t_2 - 17 t_1 t_2 - 39 t_1^2 t_2 - 5 t_1^3 t_2 + 
 5 t_1 t_2^2 \\&+& 39 t_1^2 t_2^2 + 17 t_1^3 t_2^2 - t_1^4 t_2^2- t_1 t_2^3 - 
 9 t_1^2 t_2^3 - 9 t_1^3 t_2^3 - t_1^4 t_2^3\,.
\eeqrn
\noindent{\bf Step (iii):} We next compute the numerator $N_{\bf R}(t_1,t_2;z_1,z_2,z_3)$ of $G_{\bf R}$ by provisionally assuming that the only non-vanishing coefficients correspond to the monomials appearing in the numerator of $E_{\bf R}(t_1,t_2)$. After using the eigenvalue equation (\ref{eq:sch}) to figure out the real characters needed, we get
\beqrn
N_{\bf R}&=&1 + t_1^9 -t_2^2+ t_1 t_2^4 -  t_1^{10} t_2^2+ t_1^{10} t_2^4 + 3(t_1 + t_1^8 + t_1^2 t_2^4 + t_1^9 t_2^4) + n_1 (t_1 t_2 + t_1^9 t_2^3)\\&+& n_2(t_1 t_2^2 + t_1^9 t_2^2) -z_2 (t_1^9 t_2 + t_1 t_2^3)- n_3(t_1^2 + t_1^7 + t_1^3 t_2^4 + t_1^8 t_2^4) + n_4(t_1^2 t_2 + t_1^8 t_2^3)\\&+& n_5(t_1^2 t_2^2 + t_1^8 t_2^2) -2 z_2 (t_1^8 t_2 + t_1^2 t_2^3) +  n_6(t_1^3 + t_1^6 + t_1^4 t_2^4 + t_1^7 t_2^4) + n_7(t_1^3 t_2 + t_1^7 t_2^3)\\&+& n_8(t_1^3 t_2^2 + t_1^7 t_2^2) + n_9(t_1^7 t_2 + t_1^3 t_2^3) + n_{10}(t_1^4 + t_1^5 + t_1^5 t_2^4 + t_1^6 t_2^4)  + n_{11}(t_1^4 t_2 + t_1^6 t_2^3)\\&+&  n_{12}(t_1^4 t_2^2 + t_1^6 t_2^2) +n_{13} (t_1^6 t_2 + t_1^4 t_2^3)  +  n_{14} (t_1^5 t_2 + t_1^5 t_2^3)+ n_{15} t_1^5 t_2^2\,,
\eeqrn
where the coefficients are
\beqrn
n_1&=&z_2-z_1^2-z_3^2\,,\\
n_2&=&2 z_1 z_3-4\,,\\
n_3&=&z_2^2-6\,,\\
n_4&=&2 z_1 z_2 z_3-3 z_1^2+2 z_2-3 z_3^2\,,\\
n_5&=&z_2^2-z_1^2 z_2+6 z_1 z_3-z_2 z_3^2-9\,,\\
n_6&=&z_1^2 z_2-3 z_2^2-2 z_1 z_3+z_2 z_3^2+10\,,\\
n_7&=&8 z_1 z_2 z_3-5 z_1^2+2 z_2-z_2^3-z_1^3 z_3-5 z_3^2-z_1 z_3^3\,,\\
n_8&=&z_1^4-5 z_1^2 z_2+4 z_2^2+14 z_1 z_3-5 z_2 z_3^2+z_3^4-16\,,\\
n_9&=&z_2^3-z_1^2-2 z_2-z_3^2\,,\\
n_{10}&=&2 z_1^2 z_2-4 z_2^2-2 z_1 z_3-z_1^2 z_3^2+2 z_2 z_3^2+12\,,\\
n_{11}&=&10 z_1 z_2 z_3-6 z_1^2+2 z_2-2 z_2^3-z_1^3 z_3-6 z_3^2-z_1 z_3^3\,,\\
n_{12}&=&2 z_1 z_2^2 z_3+2 z_1^4-10 z_1^2 z_2+7 z_2^2+22 z_1 z_3-z_1^2 z_3^2-10 z_2 z_3^2+2 z_3^4-22\,,\\
n_{13}&=&4 z_1 z_2 z_3-3 z_1^2-2 z_2-z_1^2 z_2^2+2 z_2^3-3 z_3^2-z_2^2 z_3^2\,,\\
n_{14}&=&8 z_1 z_2 z_3-5 z_1^2-z_1^2 z_2^2-z_1^3 z_3-5 z_3^2+z_1^2 z_2 z_3^2-z_2^2 z_3^2-z_1 z_3^3\,,\\
n_{15}&=&4 z_1 z_2^2 z_3+3 z_1^4-12 z_1^2 z_2+8 z_2^2+24 z_1 z_3-z_1^3 z_2 z_3-12 z_2 z_3^2-z_1 z_2 z_3^3+3 z_3^4-24\,.
\eeqrn

\medskip
\noindent{\bf Step (iv):} There only remains to find out if
\beq
G_{\bf R}(t_1,t_2;z_1,z_2,z_3)=\frac{N_{\bf R}(t_1,t_2;z_1,z_2,z_3)}{D_{\bf R}(t_1,t_2;z_1,z_2,z_3)} \label{eq:gareal3}
\eeq
solves the differential equation
\bdm
(\Delta^{\bf R}_t-\Delta_z)G_{\bf R}(t_1,t_2;z_1,z_2,z_3)=0\,,
\edm
where the explicit form of $\Delta^{\bf R}_t$  is derived from (\ref{eq:eig}) in the usual way:
\bdm
\Delta^{\bf R}_t=4 t_1^2 \partial_{t_1}^2+2 t_2^2 \partial_{t_2}^2+4 t_1 t_2 \partial_{t_1} \partial_{t_2}+ 16 t_1 \partial_{t_1}+10 t_2 \partial_{t_2} .
\edm
The result of this checking is positive and we can thus conclude that (\ref{eq:gareal3}) is the generating function we were seeking for.
\section{Conclusions and outlook}
The technique for computing generating functions for characters of simple Lie algebras introduced in \cite{nfp14} has by now been used to obtain a variety of results concerning characters and weight multiplicities for rank two algebras in \cite{nfp15a} and \cite{nfp15b} and to study the case of the rank three algebra $A_3\simeq D_3$ in the present paper. These works have made patent the versatility and usefulness of the method, which enabled us to present a number of results with potential applicability in mathematics and mathematical physics. It seems, however, that the algebra considered in this paper is the highest rank one in which the formulas obtained though this approach are kept under a reasonable size. 

Thus, for instance, we have computed the generating functions for characters of the remaining algebras of rank three, $B_3$ and $C_3$, but the results are exceedingly complicated, with respectively 311 and 315 terms in the numerator, and with coefficients that in many cases are long expressions in $z$-variables, see the ancillary files. In fact, for the case of $C_3$ the time needed by Mathematica to complete step (iv) was so long that we could only check the result by verifying, in many high order examples, that the characters obtained from the generating functions do indeed satisfy the Schr\"{o}dinger equation. Therefore, for these algebras, we present only a few of the simplest results, in which the complete procedure (i)-(iv) can be readily applied. In the standard notation in which $\alpha_3$ is the root of unequal length, the generating function for the characters of the representations $R_{m_1\lambda_1}$ and $R_{m_3\lambda_3}$ of $B_3$ are, respectively
\beq
\frac{1 + t_1}{1 + t_1^6 - (t_1+t_1^5) (z_1-1) + 
 (t_1^2+t_1^4) (z_2 - z_1 + 1) - t_1^3 (z_3^2-2 z_2-2)}\label{eq:b31}
\eeq
and
\beq
\frac{1 - t_3^2}{1 + t_3^8 - (t_3+t_3^7) z_3+ (t_3^2 +t_3^6)(z_1 + z_2) - (t_3^3+t_3^5) z_1 z_3 + t_3^4 ( z_1^2+ z_3^2 - 2 z_2 -1)}\,.\label{eq:b32}
\eeq
The denominator of the generating function for the representations $R_{m_1\lambda_1+m_3\lambda_3}$ is the product of the denominators of (\ref{eq:b31}) and (\ref{eq:b32}) and the numerator is
\bdm
1 + t_1 - t_3^2 - t_1^3 t_3^2 + t_1^2 t_3^4 + t_1^3 t_3^4 + 
 t_1 (1 + t_1) t_3^2 z_1 - t_1 t_3 (1 + t_1 t_3^2)  z_3\,.
\edm

The analogous of (\ref{eq:b31}) and (\ref{eq:b32}) for $C_3$ are, respectively, given by
\beq
\frac{1}{1 + t_1^6 - (t_1+t_1^5) z_1 + (t_1^2+t_1^4) (z_2+1) - 
 t_1^3 (z_1 + z_3)} \label{eq:c31}
\eeq
and
\beq
\frac{1 - t_3^4 + t_3 z_1 - t_3^3 z_1}{1 + t_3^8 + (t_3+t_3^7) (z_1 - z_3) + (t_3^2+t_3^6) (z_2^2 - 2 z_1 z_3) +  
   (t_3^3+t_3^5) (z_1^2 - 2 z_2-1) (z_1 - z_3)+ 
   t_3^4 C}\,,\label{eq:c32}
\eeq
with $C= z_1^2 + z_1^4 - 4 z_1^2 z_2 + 2 z_2^2 + 2 z_1 z_3 + z_3^2-2$. In the case of representations $R_{m_1\lambda_1+m_3\lambda_3}$ the denominator is the product of the denominators of (\ref{eq:c31}) and (\ref{eq:c32}) and the numerator is
\beqrn
 1 - t_1^3 t_3 - t_3^4 + t_1^3 t_3^5 &+& 
 t_3 (1 + t_1^2 - t_1^3 t_3 - t_3^2 + t_1 t_3^3 + t_1^3 t_3^3) z_1 
 \\&-&t_1 t_3 (1 + t_1 t_3) (1 + t_3^2) z_2 - t_1 t_3^2 (1 + t_1 t_3) z_3 + 
 t_1 t_3^2 (t_1 + t_3) z_1^2.
\eeqrn

It looks like that it is only for special cases of this type that results of a manegeable size are to be obtained if the approach based in Calogero-Sutherland model is applied to other higher rank classical Lie algebras or to the exceptional ones.

\section*{Appendix}
We give here the the form of the numerators $N_{n_1,n_2,n_3}(t_1,t_2,t_3)$ of the generating functions for weight multiplicities for the cases $n_1+n_2+n_3\leq 2$.  The cases not explicitly written arise through the change $t_1\leftrightarrow t_3$ on the appropriate numerator.
\beqrn
N_{0,0,0} &=& 1 + 2 t_1^2 t_2 + t_1^4 t_2^2 + t_1 t_3 + t_1^3 t_2 t_3 + t_1 t_2^2 t_3 - 2 t_1^5 t_2^2 t_3 - t_1^3 t_2^3 t_3 + t_1^2 t_3^2 + 
     2 t_2 t_3^2\\ &-& 2 t_1^4 t_2 t_3^2 - 4 t_1^2 t_2^2 t_3^2+ t_1^6 t_2^2 t_3^2 - 2 t_1^4 t_2^3 t_3^2 + t_1^3 t_3^3 + t_1 t_2 t_3^3 - 
     t_1^5 t_2 t_3^3 - 2 t_1^3 t_2^2 t_3^3 - t_1 t_2^3 t_3^3  \\&+& t_1^5 t_2^3 t_3^3 + t_1^3 t_2^4 t_3^3 - 2 t_1^2 t_2 t_3^4 + t_2^2 t_3^4- 
     4 t_1^4 t_2^2 t_3^4 - 2 t_1^2 t_2^3 t_3^4 + 2 t_1^6 t_2^3 t_3^4 + t_1^4 t_2^4 t_3^4 - t_1^3 t_2 t_3^5 
     \\&-&2 t_1 t_2^2 t_3^5 + t_1^5 t_2^2 t_3^5 + t_1^3 t_2^3 t_3^5 + t_1^5 t_2^4 t_3^5 + t_1^2 t_2^2 t_3^6+ 2 t_1^4 t_2^3 t_3^6 + t_1^6 t_2^4 t_3^6\,,
     \\[10pt]
N_{1,0,0} &=& t_1 + 2 t_1^3 t_2 + t_1 t_2^2 + t_1^2 t_3 + 2 t_2 t_3 - t_1^4 t_2 t_3 - t_1^2 t_2^2 t_3 - t_1^4 t_2^3 t_3 + t_1^3 t_3^2 + 
     t_1 t_2 t_3^2 - t_1^5 t_2 t_3^2 \\&-& 3 t_1^3 t_2^2 t_3^2- t_1 t_2^3 t_3^2 - t_1^5 t_2^3 t_3^2 + t_3^3 + t_2^2 t_3^3 - 3 t_1^4 t_2^2 t_3^3 - 
     2 t_1^2 t_2^3 t_3^3 + 2 t_1^6 t_2^3 t_3^3 + t_1^4 t_2^4 t_3^3 \\&-& 2 t_1^3 t_2 t_3^4- 2 t_1 t_2^2 t_3^4 - t_1^5 t_2^2 t_3^4+ 
     t_1^5 t_2^4 t_3^4 - t_1^4 t_2 t_3^5 - 2 t_1^2 t_2^2 t_3^5 + t_1^6 t_2^2 t_3^5 + t_1^4 t_2^3 t_3^5 + t_1^6 t_2^4 t_3^5 \\&+& 
     t_1^3 t_2^2 t_3^6 + 2 t_1^5 t_2^3 t_3^6 + t_1^3 t_2^4 t_3^6\,,
  \\[10pt]
N_{0,1,0} &=& t_1^2 + t_2 + t_1^4 t_2 + t_1^2 t_2^2 + t_1^3 t_3 + 2 t_1 t_2 t_3 - t_1^5 t_2 t_3 - t_1^3 t_2^2 t_3 - t_1^5 t_2^3 t_3 + t_3^2 + 
     t_2^2 t_3^2 \\&-& 4 t_1^4 t_2^2 t_3^2 - 3 t_1^2 t_2^3 t_3^2+ t_1^6 t_2^3 t_3^2 + t_1 t_3^3 - t_1 t_2^2 t_3^3 - t_1^5 t_2^2 t_3^3 + 
     t_1^5 t_2^4 t_3^3 + t_2 t_3^4 - 3 t_1^4 t_2 t_3^4 \\&-& 4 t_1^2 t_2^2 t_3^4 + t_1^6 t_2^2 t_3^4 + t_1^6 t_2^4 t_3^4 - t_1 t_2 t_3^5-
     t_1^3 t_2^2 t_3^5 - t_1 t_2^3 t_3^5 + 2 t_1^5 t_2^3 t_3^5 + t_1^3 t_2^4 t_3^5 + t_1^4 t_2^2 t_3^6 \\&+& t_1^2 t_2^3 t_3^6 + 
     t_1^6 t_2^3 t_3^6 + t_1^4 t_2^4 t_3^6\,,
  \\[10pt]
N_{2,0,0} &=& t_1^2 + 2 t_1^4 t_2 + t_1^2 t_2^2 + t_2^3 - t_1^4 t_2^3 + t_1^3 t_3 + 2 t_1 t_2 t_3 - t_1^5 t_2 t_3 - t_1^3 t_2^2 t_3 - 
     t_1^5 t_2^3 t_3 + t_1^4 t_3^2 \\&+& t_1^2 t_2 t_3^2 - t_1^6 t_2 t_3^2+ 3 t_2^2 t_3^2 - 6 t_1^4 t_2^2 t_3^2 - 4 t_1^2 t_2^3 t_3^2 + 
     2 t_1^6 t_2^3 t_3^2 - t_2^4 t_3^2 + t_1^4 t_2^4 t_3^2 + t_1 t_3^3 - t_1 t_2^2 t_3^3 \\&-&  t_1^5 t_2^2 t_3^3 + t_1^5 t_2^4 t_3^3 + 
     2 t_2 t_3^4 - 4 t_1^4 t_2 t_3^4 - 4 t_1^2 t_2^2 t_3^4 + t_1^6 t_2^2 t_3^4 - t_2^3 t_3^4 + t_1^4 t_2^3 t_3^4 + t_1^6 t_2^4 t_3^4 - 
     t_1 t_2 t_3^5 \\&-& t_1^3 t_2^2 t_3^5 - t_1 t_2^3 t_3^5 + 2 t_1^5 t_2^3 t_3^5 + t_1^3 t_2^4 t_3^5+ t_3^6 - t_1^4 t_3^6 - t_1^2 t_2 t_3^6 + 
     t_1^6 t_2 t_3^6 - 2 t_2^2 t_3^6 + 3 t_1^4 t_2^2 t_3^6 \\&+& 2 t_1^2 t_2^3 t_3^6 + t_2^4 t_3^6\,,
  \\[10pt]
N_{1,1,0} &=& t_1^3 + t_1 t_2 + t_1^5 t_2 + t_1^3 t_2^2 + t_1 t_2^3 - t_1^5 t_2^3 + t_1^4 t_3 + 2 t_1^2 t_2 t_3 - t_1^6 t_2 t_3 + 2 t_2^2 t_3 - 
     3 t_1^4 t_2^2 t_3 \\&-&  2 t_1^2 t_2^3 t_3 + t_1^6 t_2^3 t_3+ t_1 t_3^2 - 3 t_1^5 t_2^2 t_3^2 - 2 t_1^3 t_2^3 t_3^2 - t_1 t_2^4 t_3^2 + 
     t_1^5 t_2^4 t_3^2 + t_1^2 t_3^3 + 2 t_2 t_3^3 - 2 t_1^4 t_2 t_3^3 \\&-& 3 t_1^2 t_2^2 t_3^3+ t_1^6 t_2^2 t_3^3 + t_1^6 t_2^4 t_3^3 -
     2 t_1^5 t_2 t_3^4 - 3 t_1^3 t_2^2 t_3^4 - 2 t_1 t_2^3 t_3^4 + 2 t_1^5 t_2^3 t_3^4 + t_1^3 t_2^4 t_3^4 + t_3^5 - t_1^4 t_3^5\\&-& 
     2 t_1^2 t_2 t_3^5 +t_1^6 t_2 t_3^5 - t_2^2 t_3^5 + t_1^6 t_2^3 t_3^5 + t_1^4 t_2^4 t_3^5 - t_1 t_2^2 t_3^6 + 2 t_1^5 t_2^2 t_3^6 + 
     2 t_1^3 t_2^3 t_3^6 + t_1 t_2^4 t_3^6\,,
  \\[10pt]
 N_{1,0,1} &=& t_1^4 + 2 t_1^2 t_2 + t_2^2 + t_1 t_3 + t_1^3 t_2 t_3 + t_1 t_2^2 t_3 - 2 t_1^5 t_2^2 t_3 - t_1^3 t_2^3 t_3 + t_1^2 t_3^2 + 
     2 t_2 t_3^2 - 2 t_1^4 t_2 t_3^2 \\&-& 3 t_1^2 t_2^2 t_3^2 - 2 t_1^4 t_2^3 t_3^2 - t_1^2 t_2^4 t_3^2 + t_1^6 t_2^4 t_3^2 + t_1^3 t_3^3 + 
     t_1 t_2 t_3^3 - t_1^5 t_2 t_3^3 - 2 t_1^3 t_2^2 t_3^3 - t_1 t_2^3 t_3^3 + t_1^5 t_2^3 t_3^3 \\&+& t_1^3 t_2^4 t_3^3 + t_3^4- t_1^4 t_3^4 - 
     2 t_1^2 t_2 t_3^4 - 3 t_1^4 t_2^2 t_3^4 - 2 t_1^2 t_2^3 t_3^4 + 2 t_1^6 t_2^3 t_3^4 + t_1^4 t_2^4 t_3^4 - t_1^3 t_2 t_3^5 - 
     2 t_1 t_2^2 t_3^5\\ &+& t_1^5 t_2^2 t_3^5 + t_1^3 t_2^3 t_3^5 + t_1^5 t_2^4 t_3^5+ t_1^6 t_2^2 t_3^6 + 2 t_1^4 t_2^3 t_3^6 + t_1^2 t_2^4 t_3^6\,,
  \\[10pt]
N_{0,2,0} &=& t_1^4 + t_1^2 t_2 + t_1^6 t_2 + t_2^2 + t_1^2 t_2^3 - t_1^6 t_2^3 + t_1^5 t_3 + 2 t_1^3 t_2 t_3 - t_1^7 t_2 t_3 + 2 t_1 t_2^2 t_3 - 
     3 t_1^5 t_2^2 t_3 \\&-& 2 t_1^3 t_2^3 t_3 + t_1^7 t_2^3 t_3+ t_1^2 t_3^2 + t_2 t_3^2 - t_1^4 t_2 t_3^2 - t_1^2 t_2^2 t_3^2 - 
     2 t_1^6 t_2^2 t_3^2 + t_2^3 t_3^2 - 3 t_1^4 t_2^3 t_3^2 - 3 t_1^2 t_2^4 t_3^2 \\&+& 3 t_1^6 t_2^4 t_3^2 + t_1^3 t_3^3 + 2 t_1 t_2 t_3^3- 
     2 t_1^5 t_2 t_3^3 - 3 t_1^3 t_2^2 t_3^3 + t_1^7 t_2^2 t_3^3 - 2 t_1 t_2^3 t_3^3 + 2 t_1^5 t_2^3 t_3^3 + 2 t_1^3 t_2^4 t_3^3\\&-& 
     t_1^7 t_2^4 t_3^3 + t_3^4 - t_1^4 t_3^4 - t_1^2 t_2 t_3^4 - t_1^6 t_2 t_3^4 - 3 t_1^4 t_2^2 t_3^4 - 3 t_1^2 t_2^3 t_3^4 + 
     3 t_1^6 t_2^3 t_3^4 + t_1^4 t_2^4 t_3^4 + t_1 t_3^5 \\&-& t_1^5 t_3^5 - 2 t_1^3 t_2 t_3^5 + t_1^7 t_2 t_3^5 - 3 t_1 t_2^2 t_3^5 + 
     2 t_1^5 t_2^2 t_3^5 + 2 t_1^3 t_2^3 t_3^5 - t_1^7 t_2^3 t_3^5 + t_1^5 t_2^4 t_3^5 + t_2 t_3^6 \\&-& t_1^4 t_2 t_3^6- 2 t_1^2 t_2^2 t_3^6 + 
     3 t_1^6 t_2^2 t_3^6 - t_2^3 t_3^6 + 3 t_1^4 t_2^3 t_3^6 + 3 t_1^2 t_2^4 t_3^6 - 2 t_1^6 t_2^4 t_3^6 - t_1 t_2 t_3^7 + t_1^5 t_2 t_3^7 \\&+& 
     t_1^3 t_2^2 t_3^7 - t_1^7 t_2^2 t_3^7 + t_1 t_2^3 t_3^7 - t_1^5 t_2^3 t_3^7 - t_1^3 t_2^4 t_3^7 + t_1^7 t_2^4 t_3^7\,,
\eeqrn

\end{document}